\documentclass[prd,onecolumn]{revtex4}
\usepackage{epsfig,amsmath,amssymb,mathtools,amscd}
\usepackage[active]{srcltx}
\usepackage[linkcolor=red]{hyperref} 
\usepackage{dsfont}
\usepackage{subfigure}

\def\ie{{i.~e.}}
\mathchardef\minus="002D

\def\<{\langle}\def\>{\rangle}
 \def\ket#1{| #1 \rangle} 
\def\bra#1{\langle #1 |}

\def\Reals{\mathbb{R}}

\def\L2{{\mathcal L}_2}

\def\d#1 {\mathop{\!\! \mathrm{d}#1}\,}
\def\df#1#2 {\!\!\frac{\mathop{\mathrm{d}#1}}{#2}\,}

\def\bvec#1{\mathbf{#1}}
\def\bk{\bvec k}
\def\bh{\bvec h}

\def\bn{\bvec n}

\def\bg{\bvec g}

\def\bn{\bvec n}

\def\d{\operatorname{d}}
\def\Exp{\operatorname{Exp}}

\begin{document}
\title{Free quantum field theory from quantum cellular automata: \\
derivation of Weyl, Dirac and Maxwell quantum cellular
  automata\footnote{Work presented (together with Ref.~\cite{Bisio2015}) at the conference \emph{Quantum
      Theory: from Problems to Advances}, held on 9-12 June 2014 at at
    Linnaeus University, Växjö University, Sweden.}}

\author{Alessandro \surname{Bisio}} \email[]{alessandro.bisio@unipv.it} \affiliation{QUIT group, Dipartimento di
  Fisica, Universit\`{a} degli Studi di Pavia, via Bassi 6, 27100 Pavia, Italy}
\affiliation{Istituto Nazionale di Fisica Nucleare, Gruppo IV, via Bassi 6, 27100 Pavia, Italy}

\author{Giacomo Mauro \surname{D'Ariano}} \email[]{dariano@unipv.it}
\affiliation{QUIT group, Dipartimento di Fisica, Universit\`{a} degli
   Studi di Pavia, via Bassi 6, 27100
  Pavia, Italy}
\affiliation{Istituto Nazionale di
  Fisica  Nucleare, Gruppo IV, via Bassi 6, 27100
  Pavia, Italy}

\author{Paolo \surname{Perinotti}}
\email[]{paolo.perinotti@unipv.it} 
\affiliation{QUIT group, Dipartimento di Fisica, Universit\`{a} degli
   Studi di Pavia, via Bassi 6, 27100
  Pavia, Italy}
\affiliation{Istituto Nazionale di
  Fisica  Nucleare, Gruppo IV, via Bassi 6, 27100
  Pavia, Italy}

\author{Alessandro \surname{Tosini}}
\email[]{paolo.perinotti@unipv.it} 
\affiliation{QUIT group, Dipartimento di Fisica, Universit\`{a} degli
   Studi di Pavia, via Bassi 6, 27100
  Pavia, Italy}
\affiliation{Istituto Nazionale di
  Fisica  Nucleare, Gruppo IV, via Bassi 6, 27100
  Pavia, Italy}

\begin{abstract} 
  After leading to a new axiomatic derivation of quantum theory, the
  new informational paradigm is entering the domain of quantum field
  theory, suggesting a quantum automata framework that can be regarded
  as an extension of quantum field theory to including an hypothetical
  Planck scale, and with the usual quantum field theory recovered in
  the relativistic limit of small wave-vectors. Being derived from
  simple principles (linearity, unitarity, locality, homogeneity,
  isotropy, and minimality of dimension), the automata theory is {\em
    quantum ab-initio}, and does not assume Lorentz covariance and
  mechanical notions. Being discrete it can describe localized states
  and measurements (unmanageable by quantum field theory), solving all
  the issues plaguing field theory originated from the
  continuum. These features make the theory an ideal framework for
  quantum gravity, with relativistic covariance and space-time
  emergent solely from the interactions, and not assumed a priori.

The paper presents a synthetic derivation of the automata theory, showing how from the principles lead to a description in terms of a quantum automaton over a Cayley graph of a group. Restricting to Abelian groups we show how the automata recover the Weyl, Dirac and Maxwell dynamics in the relativistic limit. We conclude with some new routes about the more general scenario of non-Abelian Cayley graphs.
\end{abstract}

\maketitle

\section{Introduction}
The field of Quantum Information has been intimately linked to quantum
foundations since the very beginning, when it shed new light on {\em
  entanglement} and {\em nonlocality}. The study of quantum protocols,
like tomography, teleportation, cloning, eventually provided a
significant boost to the reformulation of QT, leading to the idea that
QT could have been regarded as a theory of information, namely
asserting basic properties of information-processing, such as the
possibility or impossibility to carry out certain tasks by
manipulating physical systems. In this scenario the work
\cite{hardy2001quantum} reopened the debate about the operational
axiomatizations
\cite{fuchs2002quantum,dariano114,d2010probabilistic,quit-purification,dakic2009quantum,masanes2011derivation}
ultimately leading to a complete derivation of finite dimensional QT
from informational principles \cite{quit-derivation}.

QT as such, however, is a general theory of systems, and does not
account for physical objects and mechanical notions. This motivates to
push the informational framework forward, to derive also quantum field
theory (QFT) from principles of information-theoretic nature. This
follow-up is also motivated by the the role that information is
playing in theoretical physics at the fundamental level of quantum
gravity and Planck-scale, e.~g. in the holographic-principle and the
ultraviolet cutoffs, implying an upper bound to the amount of
information that can be stored in a finite space volume
\cite{bekenstein1973black,hawking1975particle,PhysRevLett.90.121302}.
Imposing an {\em in-principle} upper bound to the information density,
forces us to replace continuous quantum fields with a denumerable set
of finite dimensional quantum systems, namely a quantum cellular
automaton (QCA)~\cite{feynman1982simulating} representing the unitary
evolution of quantum systems in local interaction
\cite{grossing1988quantum,aharonov1993quantum,ambainis2001one}. Thus
the QCA becomes a systematic way of exploring the informational
paradigm of Wheeler \cite{wheeler1982computer}.

After the first works
\cite{bialynicki1994weyl,meyer1996quantum,Yepez:2006p4406} recovering
relativistic quantum dynamics from QCAs, the possibility of deriving
QFT from informational principles only has been proposed in several
papers
\cite{darianopirsa,darianovaxjo2010,darianovaxjo2011,darianovaxjo2012,mauro2012quantum,dariano-asl,darianosaggiatore,PhysRevA.90.062106}. Later
other papers addressed quantum field theory in the QCA framework~
\cite{arrighi2014dirac,farrelly2013discrete}, and further derivations
from principles followed
\cite{Bisio2015244,bisio2014quantum,bisio2013dirac}, whereas a
preliminary study of Lorentz covariance distortion at the Planck scale
was presented in Ref. \cite{bibeau2013doubly}. In order be a valid
description of dynamics, the QCA model must recover the usual QFT in
the relativistic limit of wave-vectors much smaller than Planck's one
\cite{Bisio2015244,PhysRevA.90.062106,bisio2014quantum}, namely in the
limit where discreteness cannot be probed.  On the other hand, the
automaton field theory exhibit a very different behavior in the
ultra-relativistic regime of Planckian wave-vectors, where it breaks
the usual continuum symmetries that is fully recovered only in the
relativistic limit.

In this paper we review the derivation from principles of
Refs.~\cite{PhysRevA.90.062106,bisio2014quantum}, showing it leads to
a QCA on a \emph{Cayley graph} of a group.  Focusing on the simplest
case of Abelian Cayley graphs we show how the automata recover the
Weyl, Dirac and Maxwell dynamics in the relativistic limit. We
conclude with some remarks in relation to the more general scenario of
QCAs on non-Abelian Cayley graphs, where in the virtually-Abelian case
one can apply a \emph{tiling} procedure \cite{erba}, where a finite
number of sites are regrouped in a single higher-dimensional cell.

\section{QCAs with symmetries}\label{s:assumptions}

A QCA gives the evolution of a denumerable set $G$ of cells, each one
corresponding to a quantum system.  In our framework (see
Refs.~\cite{Bisio2015244,PhysRevA.90.062106}) we are interested in
exploring the possibility of an automaton description of free QFT and
thus assume the quantum systems inf $G$ to correspond to quantum
fields. Moreover we require that the amount of information in a finite
number of cells must be finite, and this leads to consider \emph{Fermionic}
modes. In Section \ref{s:maxwell}, based on
Ref.~\cite{bisio2014quantum}, we see how \emph{Bosonic} systems can be
recovered in this scenario as an approximation of many Fermionic ones.
The relation between Fermionic modes and finite-dimensional quantum
systems, say \emph{qubits}, is studied in the literature, and the two theories were proved to be computationally equivalent
\cite{Bravyi2002210}. On the other hand the quantum theory of qubits and the quantum theory of Fermions are different, mostly in the notion of what are local transformations \cite{doi:10.1142/S0217751X14300257,0295-5075-107-2-20009},
with local Fermionic operations mapped into nonlocal qubits ones and vice versa. 

From now on each cell of $G$ will host a Fermionic field
operator $\psi_{g,l}$, obeying the canonical anti-commutation relations
\begin{equation}
\{\psi_{g,l},\psi_{g',l'}\}=0,\quad\{\psi_{g,l},\psi^\dag_{g',l'}\}=\delta_{g,g'}\delta_{l,l'},
\end{equation}
where $l=1,\ldots, s_g$, $s_g$ denoting the number of field components
at each site $g\in G$.  The evolution occurs in discrete identical
steps, and in each one every cell interacts with the others. The
construction of the one-step update rule is based on the following
assumptions \cite{PhysRevA.90.062106} on the systems interaction: 1)
linearity 2) locality, 3) homogeneity, 4) unitarity and 5)
isotropy. Notice that these constraints regard both the structure of
the graph made by the interacting systems and the algebraic properties
of the map providing the update rule of the field on the graph. On one
hand our assumptions provide the bare set $G$ with a specific
structure---say an \emph{arc-transitive Cayley graph}. On the other
hand the evolution map has to be a unitary operator covariant with
respect to the symmetries of the above graph.

For convenience of the reader we remind the definition of \emph{Cayley
  graph}. Given a group $G$ and a set $S$ of generators of the group,
the Cayley graph $\Gamma(G,S)$ is defined as the colored directed
graph having vertex set $G$, edge set $\{(g,gh);g\in G, h\in S\}$, and
a color assigned to each generator $h\in S$. Notice that a Cayley
graph is \emph{regular}---i.e.~each vertex has the same degree---and
\emph{vertex-transitive}---i.e.~all sites are equivalent, in the sense that
the graph automorphism group acts transitively upon its
vertices. The Cayley graphs of a group $G$ are in one to one
correspondence with its finite presentations, with $\Gamma(G,S)$
corresponding to the presentation $\<S|R\>$, where $S$ is the generator set and
$R$ is the relator set, containing elementary closed paths on the graph. We finally
remind that a Cayley graph is said \emph{arc-transitive} when its
group of automorphisms acts transitively not only on its vertices but
also on its directed edges.

We can now analyse the consequences of the aforementioned assumptions
on the interacting systems in $G$. The \emph{linearity} prescription
means that the interaction of the field at sites $g$ and $g'$ is given
in terms of a $s_{g}\times s_g'$ {\em transition matrix} $A_{gg'}$
while the \emph{locality} assumption states that any site $g$
interacts with a finite number of sites, namely $S_g$ is finite for
every $g$. Accordingly, we denote by
$S_g$ the set of systems with $A_{gg'}\neq 0$ and we assume that
$|S_g|\leq k<\infty$ for every $g$. The update rule of the field at
site $g$ is then given by the linear operator
\begin{equation}
\psi_{g}(t+1)=\sum_{g'\in S_g} A_{gg'}\psi_{g'}(t).
\end{equation}

The \emph{homogeneity} requirement states that all the sites $g\in G$
are equivalent in the following sense: i) The cardinality $|S_g|$ is
independent of $g$ and the set of transition matrices is the same for
every $g$, namely for every $g$ one has $\{A_{gg'}\}_{g'\in S_g}=\{A_h\}_{h\in S_+}$ for some set $S$ with $|S_g|=|S_0|$. If $gh=g'$, we formally write $g=g'h^{-1}$. If we collect the elements $h^{-1}$, we have the set $S_1:=\{h^{-1}|h\in S_0\}$. Finally, we define $S:=S_0\cup S_1$. Notice that due to homogeneity the graph having the elements of $G$ as vertices and the elements of $S$ as edges, with $h^{\pm1}$ connecting $g$ to $g'$ whenever $g'=gh^{\pm1}$, is regular since $|S|$ is independent of $g\in G$; ii) Closed paths are the same from any site $g\in G$, namely if a string of transitions $h_1h_2\dots{h_N}$, $h_i\in S$ is such that $gh_1h_2\dots{h_N}=g$ for some $g$ then $g'h_1h_2\dots h_N=g'$ for every $g'\in G$.

Now, one can check that the graph $\Gamma(G,S)$ having the elements of $G$ as vertices, the couples $(g,gh)$ as edges, and the edges colored with $|S|$ colors, one for each label ${h\in S}$, represents the Cayley graph of a finitely presented group with generator set $S$ and relators set $R$ made of strings of elements of $S$ corresponding to closed paths. The proof goes as follows. First, one notice that due to homogeneity either the graph $\Gamma(G,S)$ is
connected, or it consists of disconnected copies of the same connected graph. Since in the last case we will end up with many identical copies of the same automaton we will assume without loss of generality that the graph $\Gamma(G,S)$ is connected.  Second, if we define the free group $F$ of words with letters in $S$, and the free subgroup $H$ generated by words in $R$, it is easy to check that $H$ is normal in $F$, indeed by homogeneity we know that if $r=h_1h_2\dots{h_N}\in R$ then $gr=g$, and then $grg^{-1}\in R$, for every $g\in F$. We can finally take the group $G'=F/N$ that has Cayley
graph $\Gamma(G',S)=\Gamma(G,S)$ by construction, proving that $G=G'$.

After a convenient relabeling $A_h\to A_{h^{-1}}$, the automaton can then be represented by an operator over the Hilbert space $\ell^2(G)\otimes \mathbb C^s$
\begin{equation}\label{eq:noniso}
  A=\sum_{h\in S_0} T_h\otimes A_h,
\end{equation}
where $T$ is the representation of $G$ on $\ell^2(G)$,
$T_g|g'\>=|g'g^{-1}\>$.

A fist instance of the \emph{isotropy } constraint is that if the transition from $g$ to $g'$ is possible, then also that from $g'$ to $g$ is possible, namely if $A_{gg'}\neq 0$ then $A_{g'g}\neq0$. This implies that actually, to every $h\in S$ corresponds a non-null transition matrix $A_h$. This allows us to rewrite equation \ref{eq:noniso} as
\begin{equation}
  A=\sum_{h\in S} T_h\otimes A_h.
  \label{eq:automaton}
\end{equation}
Notice that the set $S$ can be split in many ways as $S=S_+\cup S_-\cup \{e\}$, with $S_-$ the set of inverses of the elements of $S_+$, and $e$ the identity in $G$ that appears only in the presence of self-interaction.
The notion of \emph{isotropy}, saying that ``any direction on $\Gamma(G,S)$ is equivalent'', is translated in mathematical terms requiring that there exists a decomposition of $S=S_-\cup S_+\cup\{e\}$, and a faithful representation $U$ over $\mathbb{C}^s$ of a group $L$ of graph automorphisms, transitive over $S_+$, such that
one has the covariance condition
\begin{equation}\label{eq:covW}
A=\sum_{h\in S} T_h\otimes A_h=\sum_{h\in S} T_{l(h)}\otimes U_lA_hU^\dag_l,\quad \forall l\in L.
\end{equation}
Notice that as a consequence of this assumption the Cayley graph $\Gamma(G,S)$ is arc-transitive.

A covariant automaton of the form \eqref{eq:covW} describes the free
evolution of a field by a quantum algorithm with finite algorithmic
complexity, and with homogeneity and isotropy corresponding to the universality of the law given by the algorithm. 

As a consequence of the assumptions, the \emph{unitarity} condition---imposing that the operator $A$ is unitary---is given by  
\begin{align}\label{eq:unitarity}
  \sum_{h\in S}A^\dag_h A_h=\sum_{h\in S}A_h A^\dag_h=I_s,\quad
  \sum_{\shortstack{$\scriptstyle h,h'\in S$\\ $\scriptstyle
      h^{-1}h'=h''$}} A^\dag_h A_{h'}=\sum_{\shortstack{$\scriptstyle
      h,h'\in S$\\ $\scriptstyle h'h^{-1}=h''$}} A_{h'} A^\dag_{h}=0
\end{align}
in terms of the transition matrices $A_h$.

\section{QCAs and the emergent spacetime}\label{s:spacetime}

In the previous Section we have seen how our assumptions lead to a
model of evolution on a discrete computational space endowed with the
structure of Cayley graph. The usual dynamics on continuous spacetime
is expected to emerge as an effective description that holds in the
regimes where the discrete scale cannot be probed.  

Within this perspective space and time are not on an equal footing,
the space emerges from the structure of the graph while the time
variable comes from the computational steps of the automaton. This
means that from the automaton it emerges a spacetime $M$ in a given
reference frame with a fixed time direction, that is $M$ has the
Cartesian product structure $M=X\times T$, with $T$ the one
dimensional manifold corresponding to time (clearly diffeomorphic to
the real line) and the $X$ the (generally $n$-dimensional) manifold
representing space\footnote{The spacetime manifold $M$ is here
  introduced in a fixed reference frame. The notion of change of
  reference frame based on the invariance of the QCA dynamics has been
  the subject of the works \cite{bibeau2013doubly,lrntz3d}.}. The steps
of the automaton evolution can be represented as a totally ordered set
of points $t_1,t_2,\ldots$ with the metric
$d_t(t_i,t_j)=|j-i|$. Similarly on the graph we take the metric $d_x$
induced by the word-counting on the Cayley graph.

An admissible candidate for emerging time manifold is a
one-dimensional manifold $T$ with metric $\mu_t$ such that there
exists an embedding $f_t$ mapping the discrete steps of the QCA into
points of $T$ that is isometric
$\mu_t(f_t(t_i),f_t(t_j))=d_t(t_i,t_j)$. We notice that this not
single out a unique metric $\mu_t$ but a whole class of metrics. This
freedom comes from the fact that the geometric structure between two
discrete points $f_t(t_i)$ and $f_t(t_j)$ is unphysical in this
scenario. 

The identification of an emerging spatial manifold is generally more
involved because in dimension higher than one the isometric embedding
of a discrete graph in a continuous manifold is usually impossible. A
possible way out is to relax the assumption of isometric embedding by
allowing the embedding to be only \emph{quasi-isometric}.
Given two metric spaces $(M_1,d_1)$ and $(M_2,d_2)$, with $d_1$ and
$d_2$ the metric of the two spaces, a map
$f:(M_1,d_1)\rightarrow (M_2,d_2)$ is a \emph{quasi-isometry} if there
exist two constants $A\geq 1$, and $B\geq 0$, such that
\begin{equation}
\forall g_1,g_2\in M_1,\qquad d_1(g_1,g_2)/A-B\leq d_2(f(g_1),f(g_2))\leq A d_1(g_1,g_2)+B.
\end{equation}
Therefore, given a Cayley graph $\Gamma$ with word
metric $d_x$, the an admissible emerging space is a manifold $(X,d_x)$
quasi-isometric to $(\Gamma,d_x)$ via a map $f_x$.  

\begin{figure}[h!]
  \begin{center}
    \includegraphics[width=.8\textwidth]{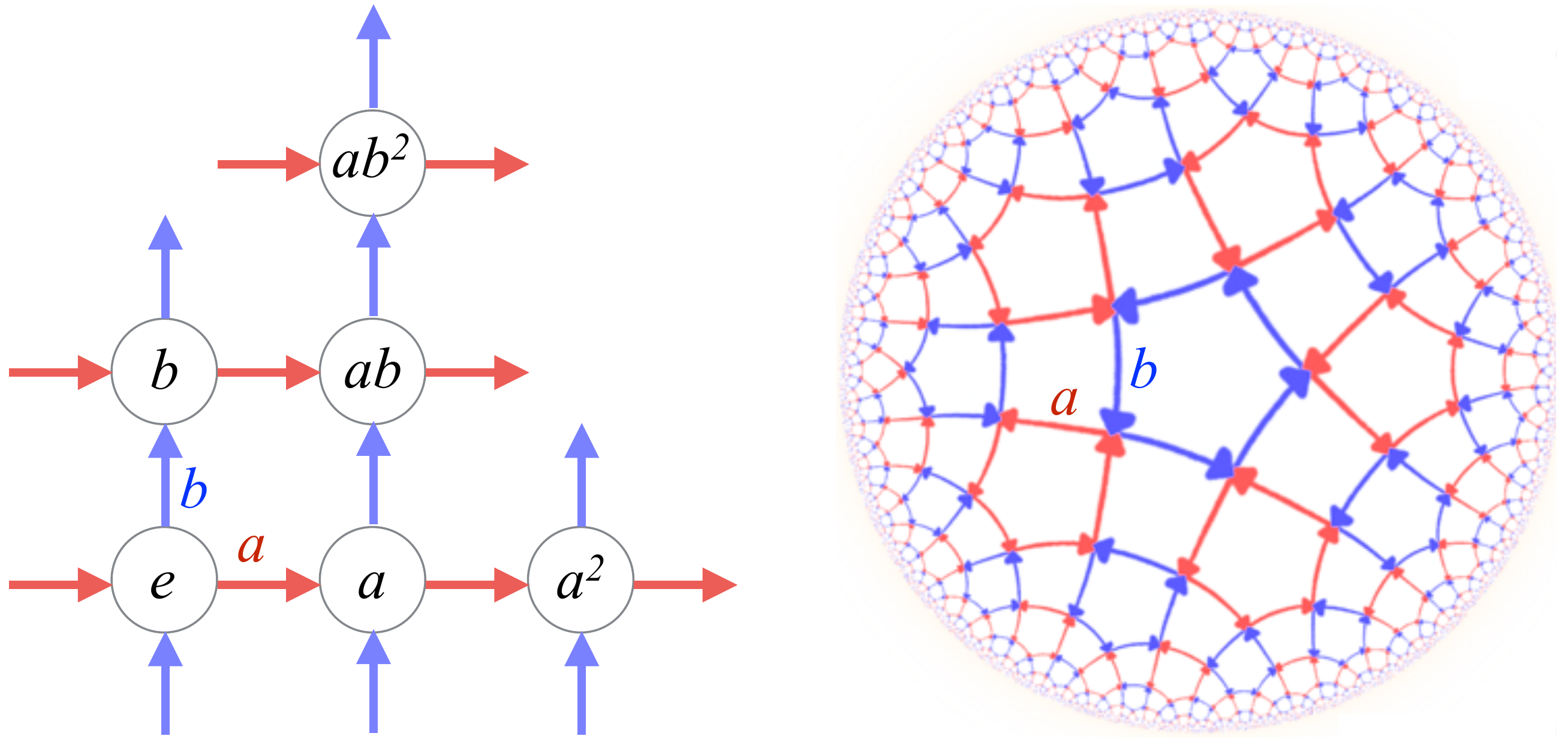}
    \caption{(colors online).  Given a group $G$ and a set $S$ of
      generators, the Cayley graph $\Gamma(G,S)$ is defined as the
      colored directed graph having set of nodes $G$, set of edges
      $\{(g,gh); g \in G, h \in S\}$, and a color assigned to each
      generator $h \in S$. {\bf Left:} the Cayley graph of the Abelian
      group $\mathbb{Z}^2$ with presentation
      $\mathbb{Z}^2=\langle a,b|aba^{-1}b^{-1}\rangle$, where a and b
        are two commuting generators. {\bf Right:} the Cayley graph of
        the non-Abelian group $G=\langle a,b|a^5,b^5,(ab)^2\rangle$.
        The Abelian-group graph is embedded into the Euclidean space
        $\mathbb{R}^2$, the non-Abelian $G$ into the Hyperbolic
        $\mathbb{H}_2$ with (negative) curvature.  }
    \label{fig:cayley}
  \end{center}
\end{figure}

The characterization of the class of metric spaces which are
quasi-isometric to a Cayley graph of a group $G$ are the subject of
\emph{geometric group theory} \cite{harpe}. It has been proved that
the quasi-isometric class is an invariant of the group, \ie it does
not depend on the group presentations (which instead correspond to
different Cayley graphs). Another basic result
\cite{gromov1984infinite} of geometric group theory is that if a group
$G$ has a Cayley graph $\Gamma$ that is quasi-isometric to the
Euclidean space $R^n$ then $G$ is \emph{virtually-Abelian}, namely it
has an Abelian subgroup $G'\subset G$ isomorphic to $\mathbb{Z}^n$ of
finite index (with a finite number of cosets). In general non-Abelian
groups are quasi-isometric to curved manifolds (see
Fig. \ref{fig:cayley}).

The presented setting of QCAs on Cayley graph can lead to field
dynamics on both flat spacetime and spacetime with curvature. In the
next Section we will focus on the flat case.

\section{QCAs on Abelian groups and the small wave-vector limit}\label{s:Abelian}
In this Section we restrict to the specific subclass of automata whose
group $G$ is quasi-isometrically embeddable in the Euclidean space,
which is then virtually-Abelian. We also assume that the
representation of the isotropy group $L$ in \eqref{eq:covW} induced by
the embedding is orthogonal, which implies that the graph neighborhood
is embedded in a sphere. In words, we want homogeneity and isotropy to
hold locally also in the embedding space. Our present analysis focus
on the Abelian groups $\mathbb Z^d$ whose Cayley graphs satisfying the
isotropic embedding in the Euclidean space $\mathbb R^d$ are the
Bravais lattices. The more general scenario of virtually-Abelian
groups is discussed in Section \ref{s:future}.

In the Abelian case (and also in the virtually-Abelian case as we will
discuss in Section \ref{s:future}) it is possible to describe the
automaton in the wave-vector space. Since the group is Abelian we
label the group elements by vectors $\bg\in\mathbb{Z}^d$, and use the
additive notation for the group composition, whereas the unitary
representation of $\mathbb{Z}^d$ on $\ell_2(\mathbb{Z}^d)$ (see
Eq.~\eqref{eq:automaton}) is expressed as
\begin{equation}
T_{\bh}|\bvec g\>=|\bh+\bg\>.
\end{equation}
Being the group Abelian, we can Fourier transform, and the operator
$A$ can be easily block-diagonalized in the wave-vector $\bk$
representation as follows
\begin{align}
  \label{eq:weylautomata} 
  A = \int_B\operatorname d^3 \! \bk  \,  |{\bk}\>\< {\bk}| \otimes
  A_{\bk},\qquad A_\bk:=\sum_{\bh\in S}\bk\,e^{-i\bk\cdot\bh}A_\bh,
\end{align}
with $A_\bk$ unitary for every $\bk\in B$, and the vectors $|{\bk}\>$
given by
\begin{equation}
|\bk\>:=\frac{1}{\sqrt{|B|}}\sum_{\bg\in G}e^{i\bk\cdot\bg}|\bg\>.
\end{equation}
Notice that due to the discreteness of the lattice the automaton is
band-limited in $\bk$ with $B$ denoting the first Brillouin zone.

The spectrum $\{e^{i\omega^{(i)}_\bk}\}$ of the operator $ A_\bk$, or
more precisely its {\em dispersion relation} that is the
expression of the phases $\omega^{(i)}_\bk$ as functions of $\bk$,
plays a crucial role in the analysis of the automaton dynamics. Indeed
the speed of the wave-front of a plane wave with wave-vector $\bk$ is
given by the {\em phase-velocity} $\omega^{(i)}_\bk/|\bk|$, while the
speed of propagation of a narrow-band state having wave-vector $\bk$
peaked around the value $\bk_0$ is given by the {\em group velocity}
at $\bk_0$, namely the gradient of the function $\omega^{(i)}_\bk$
evaluated at $\bk_0$.

\subsection{The small wave-vector limit}\label{ss:limit}

In order be a valid microscopic description of dynamics, the QCA model
must recover the usual phenomenology of QFT at the energy scale of
the current particle physics experiments, namely the physics of the
QCA model and the one of QFT must be the same as far as we restrict to
quantum states that cannot probe the discreteness of the underlying
lattice. For this reason it is important to address a comparison
between the automaton dynamics and the dynamics dictated by the usual
QFT differential equations. Here we show how to evaluate the behaviour
of an Abelian automaton for small wave-vectors $|\bk|\ll1$, and then
discuss a possible approach to a rigorous comparison at different
frequency scales.

The physical interpretation of the limit $|\bk|\ll1$ clearly depends
on the hypotheses that we make on the order of magnitude of the QCA
lattice step and time step.  In our approach
the QCA is assumed at an hypothetical discrete Planck scale, thus
the domain $|\bk|\ll1$ should correspond to the Planckian limit of the automaton. This choice clearly allows one to encompasses the relativistic regime of high energy physics within the small wave-vector limit.

In order to obtain the relativistic limit of an automaton $A_{\bk}$ we
define its {\em interpolating Hamiltonian} $H^A(\bk)$ as the operator
satisfying the following equality
\begin{equation}
  e^{-i H_I^A(\bk)}=X_\bk.
\end{equation}
The term interpolating refers to the fact that the Hamiltonian
$H_I^A(\bk)$ generates a unitary evolution that interpolates the
discrete time determined by the automaton steps through a continuous
time $t$ as
\begin{equation}\label{eq:Hamiltonian}
  \psi(\bk,t)=e^{-i H_I^A(\bk)t}\psi(\bk,0).
\end{equation}
However the dynamics provided by the Hamiltonian at the discrete
points of the graph is exactly the automaton one.

Now, one can power expand the Hamiltonian of
Eq.~\eqref{eq:Hamiltonian} to first order in $\bk$ 
\begin{equation}\label{eq:expansion}
  H_I^A(\bk)=H_A(\bk)+\mathcal O(|\bk|^2),
\end{equation}
which corresponds to the following first-order differential equation
\begin{equation}\label{eq:diff}
  i\partial_t\psi(\bk,t)=H_A(\bk)\psi(\bk,t),
\end{equation}
for narrow-band states $\psi(\bk,t)$ peaked around some $\bk_0$ with
$|\bk_0|\ll 1$.

The Hamiltonian in Eq.~\eqref{eq:diff} describes the QCA dynamics in the limit of small wave-vectors, and in the next Sections we present QCAs having the Weyl, Dirac and Maxwell Hamiltonian in as such a limit.

In Ref.~\cite{Bisio2015244} another more quantitative approach to the QFT limit of a QCA has been presented. Suppose that some automaton $A_\bk$, (with interpolating Hamiltonian $H^A_I(\bk)$) has
the unitary $U_\bk =e^{-iH_A(\bk)}$ (see Eq.~\eqref{eq:expansion}) as
first-order approximation in $\bk$. Then one can set the comparison as
a channel discrimination problem and quantify the difference between
the two unitary evolutions with the probability of error $p_e$ in the
discrimination. This probability can be computed as a function of the
discrimination experiment parameters---for example the wave-vector and
the number of particles and the duration of the evolution--- and one
can check that for values achievable in current experiments the
automaton evolution is undistinguishable from the QFT one. This
approach allows us to provide a rigorous proof that, in the limit of input states
with vanishing wave-vector, the QCA model recovers free QFT.

\section{The Weyl automaton}\label{s:weyl}

Here we present the unique QCAs on Cayley graphs of $\mathbb{Z}^d$,
$d=3,2,1$, that satisfy all the requirements of Section
\ref{s:assumptions} and with minimal internal dimension $s$ for a
non-identical evolution (see Ref.~\cite{PhysRevA.90.062106} for the
detailed derivation).

In any space dimension the only solution for $s=1$ is the identical
QCA\footnote{This was firstly noticed by Meyer in
  Ref.~\cite{meyer1996quantum} for space dimension $d=1$.}, which
means that it is not possible to have free scalar field automata\footnote{In a more general scenario scalar field automata can be
  defined as discussed in Section \ref{s:future}.}. The minimal
internal dimension for a non-trivial evolution is then $s=2$.

Let us start from the case of dimension $d=3$ that is the most
relevant from the physical perspective. For the group $\mathbb{Z}^3$
the only Cayley graphs are the primitive cubic (PC) lattice, the body
centered cubic (BCC), and the rhombohedral. However only in the BCC
case, whose presentation of $\mathbb Z^3$ involves four vectors
$S_+=\{\bh_1,\bh_2,\bh_3,\bh_4\}$ with relator
$\bh_1+\bh_2+\bh_3+\bh_4=0$, one finds solutions satisfying all the
assumptions of Section \ref{s:assumptions}. There are only four
solutions, modulo unitary conjugation, that can be divided in two
pairs $A^\pm$ and $B^\pm$. A pair of solutions is connected to the
other pair by transposition in the canonical basis,
i.e. $A_\bk^\pm=(B_\bk^\pm)^T$. The first Brillouin zone $B$ for the
BCC lattice is defined in Cartesian coordinates as $-{\sqrt3}\pi\leq
k_i\pm k_j\leq{\sqrt3}\pi, \, i\neq j\in\{x,y,z\}$ and the solutions
in the wave-vector representation are
\begin{equation}\label{eq:weyl3d}
\begin{aligned}
& A^\pm_{\bk}=I u^{\pm}_\bk-i\boldsymbol\sigma\cdot
  \tilde\bn^{\pm}_\bk,\quad B_\bk^\pm=(A_\bk^\pm)^T,\\
&{\tilde\bn}^{\pm}_{\bk} :=
\begin{pmatrix}
s_x c_y c_z \mp c_x s_y s_z\\
\mp c_x s_y c_z - s_x c_y s_z\\
c_x c_y s_z \mp s_x s_y c_z
\end{pmatrix},\quad
u^{\pm}_{\bk} :=  c_x c_y c_z \pm s_x s_y s_z ,\\
&  c_i:=\cos(k_i/\sqrt3),\quad s_i:=\sin(k_i/\sqrt3).
\end{aligned}
\end{equation}

The matrices $A^\pm_\bk$ and $B^\pm_\bk$ have spectrum
$\{e^{-i\omega^{\pm}}_\bk,e^{i\omega^{\pm}}_\bk\}$ with dispersion
relation $\omega^{\pm}_\bk=\arccos(c_xc_yc_z\mp s_xs_ys_z)$ and
evolution governed by i) the wave-vector $\bk$; ii) the helicity
direction $\bvec n^{\pm}_\bk$; and iii) the group velocity $\bvec
v^\pm_\bk:=\nabla_\bk\omega^\pm_\bk$, which represents the speed of a
wave-packet peaked around the central wave-vector $\bk$.

The above solutions satisfy the isotropy constraint
and are then covariant with respect to the group
$L'$ of binary rotations around the coordinate axes, with the
representation of the group $L'$ on $\mathbb C^2$ given by
$\{I,i\sigma_x,i\sigma_y,i\sigma_z\}$. The group $L'$ is transitive on the four BCC generators of $S_+$. 

In dimension $d=2$, the only inequivalent Cayley graphs of
$\mathbb{Z}^2$ are the square lattice and the hexagonal lattice. Also
for $d=2$ we have solutions only on one of the possible Cayley
graphs, the square lattice, whose presentation of $\mathbb Z^2$
involves two vectors $S_+=\{\bh_1,\bh_2\}$. The first Brillouin zone $B$
in this case is given by $\sqrt{2}\pi\leq k_i\leq \sqrt{2}\pi,\,
i\in\{x,y\}$ and there are only two solutions modulo unitary conjugation,
\begin{equation}
\begin{aligned}\label{eq:weyl2d}
  &A_{\bk}=I u_\bk-i\boldsymbol\sigma\cdot\tilde\bn_\bk,\quad
  B_\bk:=A_\bk^T,\\
&\tilde\bn_{\bk} :=
\begin{pmatrix}
s_x c_y\\
c_x s_y\\
s_x s_y
\end{pmatrix},\quad
u_{\bk} :=  c_x c_y,\qquad
  c_i:=\cos(k_i/\sqrt2),\quad s_i:=\sin(k_i/\sqrt2),
\end{aligned}
\end{equation}
with dispersion relation $\omega_\bk=\arccos(c_xc_y)$.

The QCA in Eq.~\eqref{eq:weyl2d} is covariant for the cyclic
transitive group generated by the transformation that exchanges
$\bh_1$ and $\bh_2$, with representation given by the rotation by
$\pi$ around the $x$-axis. Since the isotropy group has a reducible representation, the most general automaton is actually given by $(\cos\theta I+i\sin\theta\sigma_x)A_\bk$.


Finally for $d=1$ the unique Cayley graph satisfying our requirements
for $\mathbb Z$ is the lattice $\mathbb Z$ itself, presented as the
free Abelian group on one generator $S_+=\{h\}$. From the unitarity
conditions one gets the unique solution
\begin{align}\label{eq:weyl1d}
  A_k=u_\bk I-i\boldsymbol\sigma\cdot\tilde\bn_\bk,\qquad
\tilde\bn_{\bk} :=
\begin{pmatrix}
0\\
0\\
\sin k
\end{pmatrix},\quad
u_{\bk} :=  \cos k,
\end{align}
with dispersion relation $\omega_k =k$.

We call the solutions \eqref{eq:weyl3d}, \eqref{eq:weyl2d} and
\eqref{eq:weyl1d} Weyl automata, because in the limit of small
wave-vectors of Section \ref{ss:limit} their evolution obeys
Weyl's equation in space dimension $d=3$, $d=2$ and $d=1$,
respectively. Any solution in dimension $d$ is of the form
\begin{align}
W_\bk=u_\bk I-i\boldsymbol\sigma\cdot\tilde\bn_\bk,
\end{align}
for certain $u_k$ and $\bn_k$ (see Eqs.~\eqref{eq:weyl3d},
\eqref{eq:weyl2d} and \eqref{eq:weyl1d} fro $d=3,2,1$) and has dispersion relation
\begin{align}
\omega_\bk=\arccos{u_\bk}.
\end{align}\label{eq:weyl-interpolating}
It is easily to check that the interpolating Hamiltonian is
\begin{align}
H^W_I(\bk)=\boldsymbol\sigma\cdot\bn_\bk,\qquad \bn_\bk:=
\frac{\omega_{\bk}}{\sin\omega_{\bk}}\tilde{\bn}_{\bk},
\end{align}
and by power expanding at the first order in $\bk$ one has
\begin{align}
  H_I^{W}=H_{W}(\bk)+\mathcal O(|\bk|^2), \qquad
  H_{W}(\bk)=\tfrac{1}{\sqrt{d}} \boldsymbol\sigma\cdot \bk
\end{align}
where $H_{W}(\bk)$ coincides with the usual Weyl Hamiltonian in
$d$ dimensions once the wave-vector $\bk$ is interpreted as the momentum.

\section{The Dirac automaton}\label{s:dirac}

From the previous section we know that in our framework all the
admissible QCAs with $s=2$ give the Weyl equation in the limit of
small wave-vectors. In order to get a more general dynamics---say the
Dirac one---it is then necessary to increase the internal degree of
freedom $s$. Instead of deriving the most general QCAs with $s>2$. in
Ref.~\cite{PhysRevA.90.062106} is shown how the Dirac limit is
obtained from the local coupling of two Weyl automata in any space
dimension $d=1,2,3$. Here we shortly review this result.

Starting from two arbitrary Weyl automata $W$ and $W^\prime$ in
dimension $d$ (see the solutions \eqref{eq:weyl3d}, \eqref{eq:weyl2d}
and \eqref{eq:weyl1d} in for $d=3$, $d=3$ and $d=1$, respectively),
the coupling is obtained by performing the direct-sum of their
representatives $W_{\bk}$ and ${W'_\bk}$, obtaining a QCA with $s=4$,
and introducing off-diagonal blocks $X$ and $Y$ in such a way that the
obtained matrix is unitary. The locality of the coupling implies that
the off-diagonal blocks are independent of $\bk$, namely
\begin{equation}
  D_\bk:=
  \begin{pmatrix}
    p{W_\bk}&qX\\
    rY&t{W'_\bk}
  \end{pmatrix},\qquad p,q,r,t\in\mathbb{C}.
  \label{eq:diracstart}
\end{equation}

In order to satisfy all the hypothesis of Section \ref{s:assumptions}
it is possible to show that the unique local coupling of Weyl QCAs,
modulo unitary conjugation, are
\begin{equation}\label{eq:dirac-gen}
  D_\bk:=
  \begin{pmatrix}
    n W_\bk&im\\
    im&nW_\bk
  \end{pmatrix},\qquad n,m\in\Reals^+,\quad n^2+m^2=1,
\end{equation}
which are conveniently expressed in terms of gamma matrices in the
spinorial representation as follows
\begin{align}\label{eq:dirac}
  D_\bk:= nI u_\bk-i n\gamma^0\boldsymbol\gamma\cdot\tilde\bn_\bk+im\gamma^0,
\end{align}
where the functions $u_\bk$ and $\tilde\bn_\bk$ depends on the
chosen $d$ dimension Weyl automaton $W$ in Eq.~\eqref{eq:dirac-gen}.
Notice the dispersion relation of the QCAs \eqref{eq:dirac} that is
simply given by
\begin{equation}
  \omega_\bk=\arccos[\sqrt{1-m^2}u_\bk].
\end{equation}

The QCAs in Eq.~\eqref{eq:dirac-gen} are denoted Dirac QCAs because in
the small wave-vector limit narrow-band states $\psi(\bk,t)$ with
$|\bk|\ll 1$ evolves according to the usual Dirac equation.
The interpolating Hamiltonian $H_I^{D}(\bk)$ is given by
\begin{align}
  H_I^{D}(\bk)=f(\bk)(n\gamma^0\boldsymbol\gamma\cdot\tilde\bn_\bk-m\gamma^0)
  ,\qquad f(\bk):=
  \frac{\omega_{\bk}}{\sin\omega_{\bk}},
\end{align}
that by power expanding at the first order in $\bk$ is approximated as follows
\begin{align}
  H_I^{D}(\bk)=H_{D}(\bk)+\mathcal O(|\bk|^2), \qquad
  H_{D}(\bk)=\frac n{\sqrt d}\gamma^0\boldsymbol\gamma \cdot\bk+m\gamma^0.
\end{align}
Finally, for small values of $m$, $m\ll1$, we have $n\simeq 1+\mathcal
O(m^2)$ and neglecting terms of order $\mathcal O(m^2)$ and $\mathcal
O(|\bk|^2)$
\begin{equation}
  H_I^{D}(\bk)=\frac 1{\sqrt d}\gamma^0\boldsymbol\gamma \cdot\bk+m\gamma^0+O(m^2)+O(|\bk|^2),
\end{equation}
one has the Dirac Hamiltonian with the wave-vector $\bk$ and the
parameter $m$ interpreted as momentum and mass, respectively.

It is interesting to notice that in $d=1$, modulo a permutation of the
canonical basis, the Dirac QCA corresponds to two identical and
decoupled $s=2$ automata. Each of these QCAs coincide with the one
dimensional Dirac automaton derived in Ref.~\cite{Bisio2015244}. The
last one was derived as the simplest ($s=2$) homogeneous QCA covariant
with respect to the parity and the time-reversal transformation, which
are less restrictive than isotropy that singles out the only Weyl QCA
\eqref{eq:weyl1d} in one space dimension.

\section{QCA for free electrodynamics}\label{s:maxwell}

In Sections \ref{s:weyl} and \ref{s:dirac} we showed how the
dynamics of free Fermionic fields can be derived within the QCA
framework starting from informational principles. Within this
perspective the information contained in a finite number of systems
must be finite and this is the reason why at the site of the lattices
we put Fermionic modes. We now show that the same framework can also
accomodate a QCA model for the free electromagnetic field and thus for
Bosonic quantum fields with the canonical commutation relation
recovered as approximated by many Fermionic modes. The material
presented in this Section is a review of Ref.~\cite{bisio2014quantum}
where we refer for a complete presentation.

The basic idea behind this approach is to model the photon as a
correlated pair of Fermions evolving according to the Weyl QCA
presented in Section \ref{s:weyl}. Then we show that in a suitable
regime both the free Maxwell equation in three dimensions and the
Bosonic commutation relation are recovered.  Let us then consider a
couple of two component Fermionic fields, which in the wave-vector
representation are denoted as $\psi(\bk)$ and $\varphi(\bk)$. The
evolutions of these two fields are given by
\begin{align}
  \label{eq:automa2}
  {\psi} (\bk,t+1) = W_\bk{\psi} (\bk,t).
\quad
  {\varphi} (\bk,t+1) = W_\bk^*{\varphi} (\bk,t).
\end{align}
Where the matrix $W_\bk$ can be any of the Weyl QCAs in three space
dimensions of Eq.~\eqref{eq:weyl3d}, (the whole derivation is
independent on this choice) and $W_\bk^*$ denotes the complex
conjugate matrix\footnote{Since $W_\bk$ has dimension $2$, $W_\bk$
  and $W^*_\bk$ are similar through $ W_\bk^* = \sigma_y
  W_\bk\sigma_y$.}.

We now introduce the following bilinear operators
\begin{align}
  \label{eq:prephoton}
 \bvec{G}_T(\bk,t) &:=  \bvec{G}(\bk,t) -  
\left(\frac{\bvec{n}_{\frac{\bk}{2}}}{|\bvec{n}_{\frac{\bk}{2}}|} \cdot
{\bvec{G}}(\bk,t) \right)
\frac{\bvec{n}_{\frac{\bk}{2}}}{|\bvec{n}_{\frac{\bk}{2}}|} \\
\bvec{G}(\bk,t) &:= ( G^{1}(\bk,t), G^{2}(\bk,t), G^{3}(\bk,t))^T \nonumber\\
  G^{i}(\bk,t) &:= \varphi^T (\tfrac{\bk}{2},t) \sigma^{i}  \psi(\tfrac{\bk}{2} , t)
  =\varphi^T (\bk,0) ( W_{\tfrac{\bk}{2}}^\dag  \sigma^{i}
  W_{\tfrac{\bk}{2}} )\psi(\tfrac{\bk}{2} , 0) 
\nonumber
\end{align}
with $\bn_\bk$ as in Eq.~\eqref{eq:weyl-interpolating}. By
construction the field $\bvec{G}_T(\bk,t)$ obeys
\begin{align}
  \label{eq:premaxwell}
 \bvec{n}_{\frac{\bk}{2}} \cdot \bvec{G}_T(\bk,t)  &=  0,\\
  \bvec{G}_T(\bk,t) &= \Exp(-i2\bvec{n}_{\tfrac{\bk}{2}} \cdot \bvec{J}
  t)  \bvec{G}_T(\bk,0) \label{eq:premaxwell2}
\end{align}
where we used the identity 
$ \exp (-\tfrac{i}{2}\bvec{v}\cdot \boldsymbol{\sigma}) 
\boldsymbol{\sigma} \exp (\tfrac{i}{2}\bvec{v} \cdot \boldsymbol{\sigma}) =
\Exp(-i\bvec{v} \cdot \bvec{J}) \boldsymbol{\sigma} $
where the matrix $\Exp(-i\bvec{v}\cdot\bvec{J})$ acts on $\boldsymbol{\sigma}$ regarded as a vector
and $\bvec J=(J_x, J_y,J_z)$ is the vector of angular momentum
operators.
Taking the time derivative of Eq. \eqref{eq:premaxwell2} we obtain
\begin{align}
  \label{eq:premaxwell3}
  \partial_t\bvec{G}_T(\bk,t) = 2\bvec{n}_{\tfrac{\bk}{2}} \times  \bvec{G}_T(\bk,t).
\end{align}
If $\bvec{E}_G$ and
$\bvec{B}_G$ are two Hermitian operators defined by the relation
\begin{align}
  \label{eq:electric and magnetic field}
  \bvec{E}_G:=|{\bn}_{\tfrac\bk2}|(\bvec{G}_T+\bvec{G}_T^\dag),\quad\bvec{B}_G:=i|{\bn}_{\tfrac\bk2}|(\bvec{G}_T^\dag-\bvec{G}_T),
\end{align}
then Eq. \eqref{eq:premaxwell} and Eq. \eqref{eq:premaxwell3}  
can be rewritten as 
\begin{align}
  \label{eq:maxweldistorted}
  \begin{aligned}
           2\bvec{n}_{\tfrac{\bk}{2}} \cdot \bvec{E}_G &= 0 
&
     2\bvec{n}_{\tfrac{\bk}{2}} \cdot \bvec{B}_G &= 0  
\\
    \partial_t \bvec{E}_G &= i 2\bvec{n}_{\tfrac{\bk}{2}} \times
    \bvec{B}_T(\bk,t)  
&
\partial_t \bvec{B}_G &=
    - i 2\bvec{n}_{\tfrac{\bk}{2}} \times \bvec{E}_T(\bk,t)   
  \end{aligned}
 \end{align}
 that are the free Maxwell's equation in the wave-vector space with the
 substitution $ 2\bvec{n}_{\tfrac{\bk}{2}} \to \bk$.  In the limit
 $|\bk| \ll 1$ one has $ 2\bvec{n}_{\tfrac{\bk}{2}} \sim \bk$ and the
 usual free electrodynamics is recovered.

 However the field as defined in Eqs. \eqref{eq:prephoton} and
 \eqref{eq:electric and magnetic field} does not allow to recover the
 correct Bosonic commutation relation.  As shown in
 Ref.~\cite{bisio2014quantum} the solution to this problem is to
 replace the operators $G^i$ defined in Eq. \eqref{eq:prephoton} with
 the operators $F^i$ defined as
\begin{align}
  \label{eq:photon}
F^{i}(\bk) := 
 \int \frac{ d \bvec{q}}{(2 \pi)^3}
f_{\bk}(\bvec{q})
\varphi
\left(\tfrac{\bk}{2}-\bvec{q}\right)
\sigma^{i}
\psi
 \left(\tfrac{\bk}{2}+\bvec{q}\right) 
\end{align}
where
$\int\frac{d\bvec{q}}{(2\pi)^3} |f_{\bk}(\bvec{q})|^2 =1, \forall \bk$.
In terms of $\bvec{F}(\bk)$, we can define the polarization operators 
$\gamma^i(\bk)$
of the
electromagnetic field as follows
\begin{align}
 &\gamma^i(\bk) := \bvec{u}^i_\bk\cdot\bvec{F}(\bk,0),\quad i=1,2,
  \label{eq:polarization}
\\
&\bvec{u}^i_\bk \cdot \bn_{\bk} =\bvec u^1_\bk\cdot\bvec u^2_\bk= 0,
\;
 |\bvec u^i_\bk|=1,
\;
(\bvec u^1_\bk\times\bvec u^2_\bk)\cdot\bn_\bk>0.
\end{align}
In order to avoid the technicalities of the continuum we suppose to have a discrete wave-vector space (as if the electromagnetic field were confined in a finite volume) and moreover let us assume $|
{f}_\bk(\bvec{q})|^2$ to be a constant function over a region
$\Omega_\bk$ which contains $N_\bk$ modes\footnote{This derivation can  be applied, with suitable changes, to the case in which  ${f}_\bk(\bvec{q})$ is no longer a constant function, see  Ref.~\cite{bisio2014quantum}.}, i.e. $|{f}_\bk(\bvec{q})|^2 =\tfrac{1}{N_\bk}$ if $\bvec{q}\in \Omega_\bk$ and $|{f}_\bk(\bvec{q})|^2 = 0 $ if $\bvec{q} \not\in \Omega_\bk$.  Then, for a given state $\rho$ of the field we denote by $M_{\varphi,\bvec{k}}$ (resp.  $M_{\psi,\bvec{k}}$) the mean number of type $\varphi$ (resp $\psi$) Fermion in the region $\Omega_\bk$.  One can then show that, for states such that $M_{\xi,\bvec{k}}/ N_\bk \leq
\varepsilon$ for all $\xi = \varphi, \psi$ and $\bvec{k}$ and for $\varepsilon \ll 1$ we can safely assume $ [\gamma^i
(\bvec{k}),{\gamma^j}^\dag (\bvec{k}')]_- = \delta_{i,j}
\delta_{\bvec{k},\bvec{k}'}$, i.e. the polarization operators are Bosonic operators.

\section{Future perspectives: non-Abelian Cayley
  graphs and tiling}\label{s:future}

In Section \ref{s:assumptions} we have shown how from a denumerable
set of unitarily an linearly interacting systems, the locality and
homogeneity of the interactions lead to the structure of the Cayley
graph of some group $G$, with vertices corresponding to the set of
systems and edges corresponding to couples of interacting systems.

Then, we further restricted our scenario assuming the isotropy,
the equivalence of the directions on the graph, and the Abelianity of
the group $G$. Within this perspective all possible QCAs having
minimal internal degree of freedom $s=2$ for a non trivial evolution
have been derived in Section \ref{s:weyl} and give the usual Weyl
dynamics in the limit of small wave-vector. Using the Weyl QCAs, in Sections \ref{s:dirac} and \ref{s:maxwell} we 
constructed other simple automata, recovering the Dirac and the Maxwell
equation in the small wave-vector regime.

Keeping locality and homogeneity one could study automata defined on
arbitrary Cayley graphs, relaxing the hypothesis of isotropy and
Abelianity. A very general procedure that can be defined is the
\emph{tiling}. Suppose to have a QCA $A$ with internal Hilbert space
$\mathbb{C}^s$ on the Cayley graph $\Gamma(G,S)$ of some group $G$.
Whenever a group $G$ has a subgroup $G'\subset G$ of finite index $r$
it is possible to describe the same automaton $A$ as an automaton $A'$
on a Cayley graph $\Gamma(G',S')$ of $G'$ and with bigger internal
system $\mathbb{C}^{sr}$. Intuitively the information on the cosets is
included in the internal degree of freedom and the map performing this
``inclusion'' is the unitary map $U:\ell_2(G)\otimes \mathbb
C^s\rightarrow \ell_2(G')\otimes \mathbb C^{sr}$ given by
\begin{align}
  U:=\sum_{i=1}^{r} \sum\limits_{\bg'\in G'}
  \ket{\bg'}\ket{i}\bra{c_i\bg'}\otimes I_s,
\end{align}
where $\{c_1,\ldots, c_r\}$ are the representatives of the $r$ cosets.

As a first application, the tiling allows for the wave-vector space
description of any QCA on a group $G$ quasi-isometrically embeddable in
$\mathbb{R}^n$. As already stated in Section \ref{s:spacetime} these
groups are the virtually-Abelian ones, therefore they admit a finite
index Abelian subgroup $G'\subset G$, and using the tiling a QCA over
$G$ can be regarded as a QCA on the Abelian subgroup $G'$.

A second application is the construction of QCAs starting from the
tiling of simpler automata. The main motivation for this is in the
difficulty of solving the unitarity constraint in
Eq.~\eqref{eq:unitarity} for large matrices. The easiest situation is
that of scalar QCAs where the transition matrices are simply complex
numbers. While in the Abelian case scalar QCAs with a non trivial
evolution are not admissible, in a more general scenario scalar
solutions can be found \cite{Acevedo:2008:ESQ:2011752.2011757}. Within
this perspective one could explore the emergence spinorial QCAs from
the tiling of scalar automata on non-Abelian groups.

\begin{acknowledgements}
This work has been supported in part by the Templeton Foundation under the project ID\# 43796 {\em A
  Quantum-Digital Universe}.
\end{acknowledgements}

\bibliographystyle{apsrev4-1}
\bibliography{bibliography}

\end{document}